\documentclass{aa}  
\usepackage{emptypage} 
\usepackage{graphicx}
\usepackage{placeins}
\usepackage{txfonts}

\begin{document}

   \title{Continuation of an Optical Spectroscopic Campaign of Fermi Blazar Candidates with TNG: Discovery of a New Changing-Look Blazar}


   \author{N. Álvarez Crespo,
          \inst{1,2}
          A. Domínguez,
          \inst{1,2}
          V. S. Paliya,
          \inst{3}
          M. Chamorro Cazorla,
          \inst{2,4}
          P. Sánchez Blázquez,
          \inst{2,4}
          A. Gil de Paz
          \inst{2,4}
          }

   \institute{Department of EMFTEL, Universidad Complutense de Madrid, E-28040 Madrid, Spain\\
              \email{nalvarezcrespo@ucm.es, alberto.d@ucm.es}
         \and
             Instituto de Física de Partículas y del Cosmos (IPARCOS), Universidad Complutense de Madrid, E-28040 Madrid, Spain
         \and
             Inter-University Centre for Astronomy and Astrophysics (IUCAA), SPPU Campus, 411007, Pune, India\\
              \email{vaidehi.s.paliya@gmail.com}
         \and
             Departamento de Física de la Tierra y Astrofísica, Universidad Complutense de Madrid, E-28040 Madrid, Spain
             }

   \date{Received September 23, 2024; accepted December 21, 2024}

\authorrunning{Álvarez Crespo et al.}
\titlerunning{Optical Spectroscopic Campaign of Fermi BCUs with TNG}

 
  \abstract
   {Blazars are a distinct subclass of active galactic nuclei (AGN), known for their fast variability, high polarization, and intense emission across the electromagnetic spectrum, from radio waves to gamma rays. Gamma-ray blazar candidates of uncertain type (BCU) are an ongoing challenge in gamma-ray astronomy due to difficulties in classification and redshift determination.}
   {This study continues an optical spectroscopic campaign aimed at identifying the characteristics of BCUs to improve classification and redshift estimates, particularly focusing on low-synchrotron-peak sources.}
   {We conducted a detailed analysis of optical spectroscopic data for a sample of 21 low-synchrotron-peak BCUs plus one bl lac with contradictory results in the literature, using the 3.58-m Telescopio Nazionale Galileo (TNG, La Palma, Spain).}
   {Our analysis identifies 14 out of the 21 BCUs as flat-spectrum radio quasars (FSRQs), demonstrating the effectiveness of our selection criteria. Notably, four FSRQs have redshifts exceeding 1, including 4FGL J2000.0+4214 at $z = 2.04$. Six sources are classified as bl lacs, with one of them, 4FGL J0746.5-0719, showing a featureless spectrum in this work despite previously exhibiting strong lines, suggesting it may be a changing-look blazar. One source remains classified as a BCU due to a noisy spectrum. Additionally, we observed a bl lac object, 4FGL J1054.5+2211, due to inconsistent redshift estimates in the literature, but we could not confirm any redshift due to its featureless spectrum. Our findings provide insights into the classification and redshift estimation of blazar candidates, emphasizing the need for continued spectroscopic monitoring.}
   {}

   \keywords{galaxies: active – blazars: general – gamma-ray: galaxies}

\maketitle
%

\section{Introduction}
AGNs are astrophysical sources located at the center of some galaxies, powered by the accretion of matter into a supermassive black hole (10$^6$ – 10$^9$ M$_{\odot}$). About 10\% of AGNs exhibit jets, i.e.,~highly collimated outflows of ultra-relativistic particles originating from the central engine, extending tens to hundreds of kiloparsecs. Blazars are a type of AGN with relativistic jets closely aligned with the line of sight, displaying highly variable, relativistically beamed, non-thermal emission across the entire electromagnetic spectrum \cite[see e.g.][]
{blandford1976,urry1995}. The broadband spectral energy distribution (SED) of blazars shows a characteristic double-peaked structure: a lower-energy component peaking from IR to the UV and a higher-energy bump extending from X-rays to gamma rays \cite[for more information see][and references therein]{giommi1994,begue2024}.

Blazars represent approximately 40\% of the known gamma-ray sources reported in the {\it Fermi}-Large Area Telescope catalogue \citep[4FGL-DR3,][]{fgl,4fgl-dr3}. About 30\% of the sources in the 4FGL-DR3 remain unidentified, making the association of high-energy emission with its counterpart an important observational challenge in gamma-ray astrophysics. Moreover, 22\% of the sources in the 4FGL-DR3 are classified as BCUs. These sources exhibit multiwavelength properties similar to those of blazars, such as flat radio spectra, X-ray emission, and peculiar infrared colors \citep[see][]{dabrusco2014,dabrusco2019}, but lack optical spectra to unequivocally identify their nature. Optical spectra are necessary to measure distance, which is fundamental for deriving intrinsic properties. As the largest population of gamma-ray emitters, Blazars represent the most promising counterparts for a significant fraction of BCUs. Understanding the proportion of gamma-ray emission attributed to blazars is crucial for various reasons, including placing stringent limits on the extragalactic background light, especially at $z>1$ \citep[e.g.~][]{dominguez15,abdollahi18,saldana-lopez21,dominguez2024}, identifying the sources of extragalactic neutrinos \citep[e.g.~][]{icecube2018,padovani2019,buson2022}, or constraining dark matter scenarios \citep[e.g.~][]{darkmatter2015}.

According to their optical spectra, blazars are typically divided into two classes:  FSRQs, which display strong, quasar-like emission lines, and bl lacs, which show featureless optical spectra or very weak emission lines with a rest-frame equivalent width (EW) of less than 5 Å \citep{stickel91}. The featureless nature of bl lacs makes their redshift determination challenging; in fact, 40\% of the sources classified as bl lacs in the 4FGL-DR3 still have unknown distances \citep[e.g.~][]{paiano20,dominguez2024}.

Since FSRQs are known to exhibit broad and strong emission lines, we have carefully selected a sub-sample of BCUs with multi-wavelength properties similar to FSRQs, increasing the probability of detecting broad emission lines in their optical spectra and determining their redshift. Moreover, FSRQs usually show  synchrotron peaks at lower frequencies (low synchrotron-peak blazars, LSP, $\nu^{peak}_{syn} < $ 10$^{14}$ Hz), than bl lacs (high-synchrotron-peak blazars, HSP, $\nu^{peak}_{syn} > $ 10$^{15}$ Hz). Notably, there is a small fraction of bl lacs that are LSP or intermediate-synchrotron-peak blazars \citep[ISP, 10$^{14} <$ $\nu^{peak}_{syn}< $ 10$^{15}$ Hz, ][]{4lac}.

Various approaches have been developed to effectively evaluate the physical characteristics of BCUs and categorize them. For instance, one method involves their positioning in the Wide-field Infrared Survey Explorer \citep[WISE, ][]{wise} color-color diagram, where gamma-ray emitting blazars occupy a distinct region \citep{massaroircolors}. Additionally, machine learning algorithms have been employed for classification \citep[see e.g. ][]{kang2019,xiao2023,zhu2024}. However, none of these methods provide a definitive means to ascertain the nature of BCUs without optical spectroscopic confirmation. To address this issue, several optical spectroscopic follow-up initiatives have been undertaken \citep[see e.g.][]{alvarezcrespo2016a,alvarezcrespo2016b,klindt2017,marchesini2019}.

In 2020, we started an optical spectroscopic campaign to determine the nature of BCUs using ground-based optical telescopes. 
In a previous paper, \cite{olmo2022} (hereafter referred to as Paper I), we observed 27 BCU selected from the Fermi catalogue according to their low-synchrotron peak, providing strong evidence that our criteria to select potential FSRQs among BCUs are robust,
allowing the measurement of redshifts. 
Here, we present the continuation of this campaign during cycles 2022B and 2023A using the 3.58-m telescope Telescopio Nazionale Galileo (TNG, La Palma, Spain), with the aim of determining nature and redshift of BCUs. Additionally, we observed one source classified as a bl lac in the 4FGL but with contradictory redshift values reported in the literature.

The paper is organized as follows: details on sample selection are provided in section 2, while observations and data reduction are presented in section 3. The main results are reported in section 4, with notes on individual sources in section 5. Lastly, conclusions are drawn in section 6. We adopt the following cosmological parameters: $H_0$= 70 km s$^{-1}$ Mpc$^{-1}$, $\Omega_{\Lambda}$ = 0.7, and $\Omega_{M}$ = 0.3.

\section{Blazar sample}
Here we report the continuation of our optical spectroscopic campaign to discover the nature of BCUs and determine their distances. There are approximately 900 Fermi-BCUs present in the 4FGL-DR3 that are visible from La Palma ($\delta > -$20º). We selected 27 sources based on brightness (R $<$ 22) and visibility. Observations were carried out during the 2022B and 2023A observing cycles. Our main objectives are to determine, for the first time, the spectroscopic redshift and source classification based on the rest-frame EW of emission or absorption lines.

Given that FSRQs are characterized by broad and strong emission lines, we prioritized observing sources classified as LSPs in the Fourth LAT AGN Catalogue \citep[4LAC,][]{4lac} to enhance the likelihood of detecting broad emission lines in their optical spectra. This approach is supported by the fact that nearly all FSRQs exhibit LSP SEDs. As summarized in Table\ref{table:listsour}, eight BCUs in 4LAC are classified as having LSP SEDs.

Additionally, we re-observed the counterpart associated with the gamma-ray source 4FGL J1054.5+2211 due to discrepancies in the redshift value found in the literature. Given the confusion around this source, we decided to use time from our BCU campaign to observe it again.

\section{Observations and data reduction}
Table\ref{table:listsour} lists the optical counterparts observed of all 4FGL sources as given in the 4FGL catalogue, along with their observed WISE names and 4FGL information. 
BCUs are frequently identified as candidates based on their distinctive infrared colors in the WISE color-color space. Subsequently, their optical counterpart is listed in the 4FGL. To minimize the risk of missclassification or error propagation in this process, we have chosen to report both counterpart names of the observed source.
Table\ref{table:listobs} lists the observing information. All the sources in this work were observed with the 3.58-m Telescopio Nazionale Galileo (TNG) at the Roque de Los Muchachos (La Palma, Spain) in visitor mode during the 2022B and 2023A cycles over seven nights [see Table\ref{table:listobs}]. The observations were carried out using the Device Optimized for the LOw RESolution spectrograph (DOLORES), a low-resolution spectrograph and camera installed at the Nasmyth B focus of the telescope. we used the LR-B grism, a combination of a grating and a prism designed for low-resolution spectroscopy, covering the wavelength range 3000–8400 Å. To optimize observations when seeing was not optimal, we used the 1.5'' long slit configuration oriented at the parallactic angle.

Data reduction procedures were carried out using standard routines included in the Image Reduction and Analysis Facility (IRAF) package \citep{iraf}.  Standard CCD-reduction techniques were applied,
including dark-frame subtraction and flat-field correction to account for thermal noise inherent to the CCD detector, and sky-background subtraction for each recorded frame.
For cosmic ray removal, we utilized the standard IRAF procedure \texttt{cosmicrays} from the \texttt{crutil} package. This task identifies cosmic rays in the image and replaces the affected pixels with an average of their neighboring pixels. To prevent the loss of the source within the slit and  to avoid excessive cosmic-ray contamination, we limited the maximum exposure time to 1800 seconds. Consequently, observations for each source were divided into two, three, or four shorter exposures. These individual exposures were then visually compared to one another, and any residual cosmic-ray artifacts not corrected by the IRAF procedure were manually identified and removed.
. Wavelength calibration was done by observing the spectra produced by the Ne+Hg arc lamps. We observed spectro-photometric standard stars each night to perform relative flux calibration. All spectra were corrected for atmospheric extinction with the values R = 3.1 and de-reddened using the S\&F galactic extinction reddening E(B-V) value available at the NASA/IPAC Infrared Science Archive \citep{Schlafly2011}.

We searched for absorption and emission features, requiring at least two features for an unequivocal redshift determination. When only one emission line was present, we set a redshift lower limit. This approach is based on the assumption that the single emission line is Mg II, as it is often the most prominent line in FSRQs. While this assumption may not always hold true, Mg II provides the lowest redshift among other prominent lines, making it a conservative choice for establishing a redshift lower limit. Altogether, we were able to acquire optical spectra of 22 sources.

\section{Results}
In this section, we discuss the results of all 22 sources observed with TNG during the 2022B and 2023A observing cycles, over a total of seven nights. In Table\ref{table:results} we list the observing results with details of all spectral features found for each source. Notes for each individual source are detailed in section \ref{sec:notes}. All reduced spectra are shown in Figure~\ref{fig:appendix} in the Appendix.

Out of 21 BCUs observed, 14 are FSRQs, providing strong evidence that our criteria for selecting potential FSRQs among BCUs are robust. We were able to estimate the redshift for 13 of them, while for 4FGL J2139.1+3724, we could only set a lower limit since the spectrum shows only one emission line. Out of those FSRQs, four have redshifts higher than 1, including the counterpart corresponding to 4FGL J2000.0+4214, at $z = 2.04$. 

Six sources were classified as bl lacs, and due to their lack of emission or absorption features, we could only tentatively calculate the redshift for one of them, 4FGL J1517.0+2639, at $z =$ 0.05. 
The identification for the counterpart of the source 4FGL J0128.2+4400 remains inconclusive due to its noisy spectrum ($S/N =$ 2), insufficient to determine its nature, thus we retain its classification as a BCU. 
We observed the source 4FGL J0704.7+4508 for the second time, as it was previously classified in Paper I as a featureless bl lac with an unknown redshift. Again, in this work, we did not find any emission and/or absorption features that allow a redshift determination for this bl lac.

\begin{figure}
 \includegraphics[width=9.8cm]{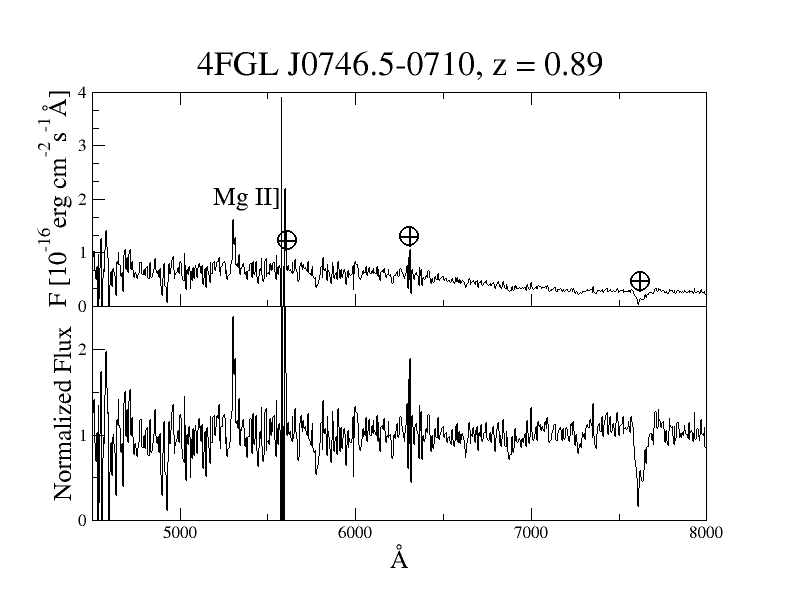}
 \includegraphics[width=9cm]{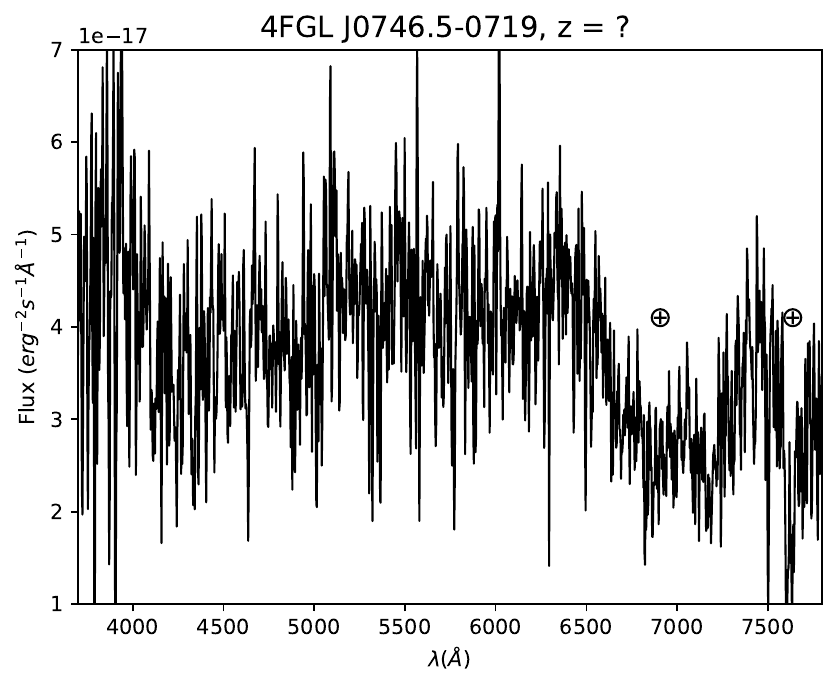}
    \caption{4FGL J0746.5-0719 spectra Top: reported in Paper I, classified as a FSRQ. Bottom: observed in this cycle. There are no visible features, thus we classify it as a bl lac. It is a changing look blazar.  Telluric lines due to atmospheric absorption are marked as crosses. }
    \label{fig:j0746}
\end{figure}

We observed the source 4FGL J0746.5-0719 in Paper I during the night of March 21, 2021, using the 3.6-m Devasthal Optical Telescope (DOT) located at the Devasthal Observatory, Nainital, India, with an exposure time of 5400 seconds and S/N = 7. In  Fig ~\ref{fig:j0746}, we report the spectra as given by Paper I in the top panel, and compare it to more recent spectra observed in this work (note that in Paper I, the authors present both the flux and the normalized flux as a preferred method for more easily representing weak features in optical spectra). Comparing both spectra, it is seen that in Paper I,  
the broad emission line Mg II (EW$_{obs}=$ 16.92 \AA) is identified, leading to a source redshift calculation of $z = 0.90$ and enabling the source to be classified as an FSRQ. However, in this work, there has been a decrease in the continuum flux and the Mg II line is not evident, so if the Mg II line was present, it should appear even more prominently. Its absence confirms that the source has undergone a transition. It is worth noting that the spectrum presented in Paper I has a comparable signal-to-noise ratio (S/N = 6) to the current one, indicating that the absence of the emission line is not attributable to noise. Interestingly, this result suggests that the source may be a changing-look blazar. We stress the importance of this discovery since there are only a few confirmed blazars that display this behavior \citep{pena2021,kang2024,paiano24}.

We re-observed the associated counterpart for the gamma-ray source classified as a bl lac, 4FGL J1054.5+2211, due to inconclusive results in the literature. Consistent with previous works, we observed a featureless bl lac and were not able to calculate its redshift.
For the source 4FGL J1754.7+3444, 4LAC reports a redshift of $z = 0.016281$ \citep{4lac}. However, we found four strong emission lines (Mg II, H$\delta$, H$\gamma$, and H$\beta$) that allow us to classify this source as an FSRQ at $z = $ 0.62.

\begin{figure}
\includegraphics[width=8.4cm]{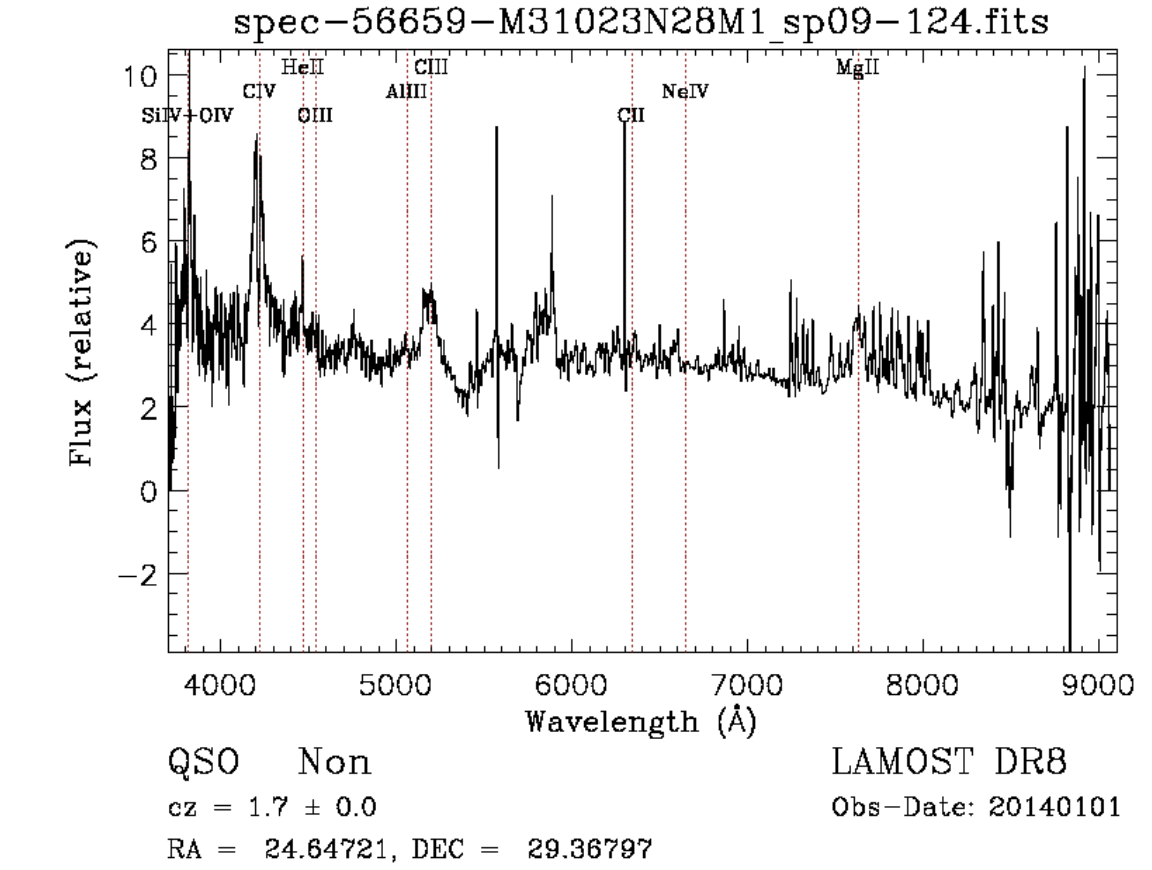}
\includegraphics[width=8.6cm]{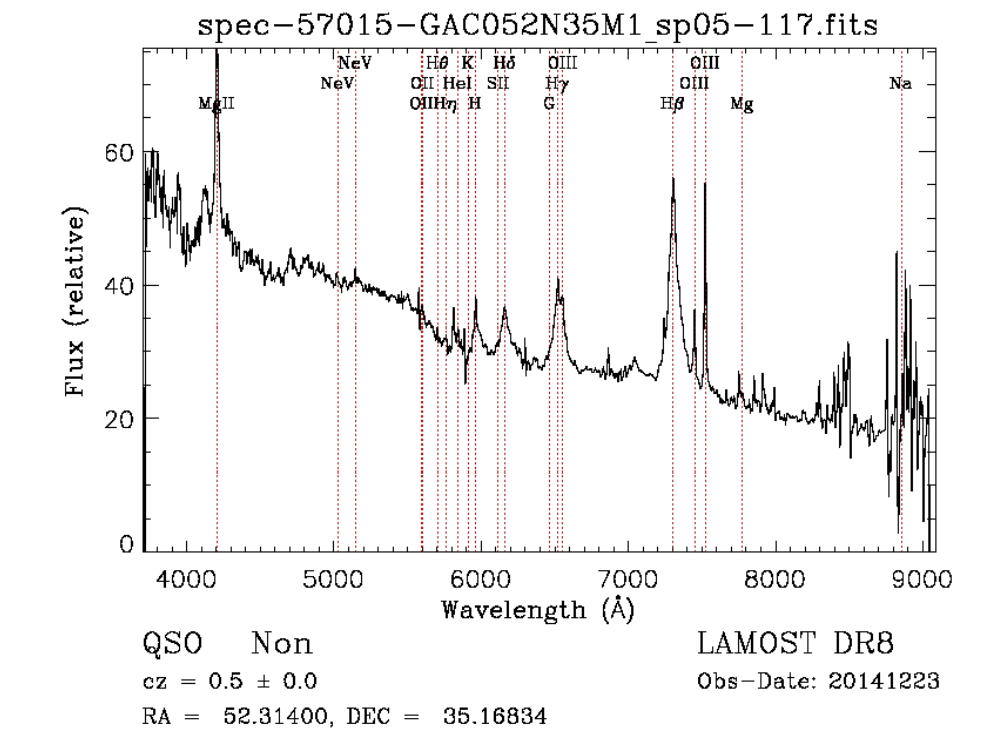}
    \caption{4FGL J0138.6+2923 and 4FGL J0328.9+3514 LAMOST LRS optical spectra.}
    \label{fig:lamost}
\end{figure}

Additionally, we checked The Large Sky Area Multi-Object Fiber Spectroscopic Telescope (LAMOST) Low-Resolution spectroscopic survey \citep[LRS,][]{lamost} General catalogue's latest public data release (DR8) to see if any of our BCUs were included. We found two sources: 4FGL J0138.6+2923 and 4FGL J0328.9+3514, as reported in Figure \ref{fig:lamost}. 4FGL J0138.6+2923 is classified as a quasar (QSO) at $z = 1.7$, the same value we found using TNG, while 4FGL J0328.9+3514 is classified as a QSO at $z = 0.5$, the same as we report here.

\section{Notes on individual sources}
\label{sec:notes}
We discuss and stress some relevant points on individual sources:

\begin{itemize}
    \item 4FGL J0128.2+4400: With such a noisy spectrum (S/N = 2), it is not possible to distinguish any feature from noise. The classification of this source is inconclusive, and it remains a BCU.
    \item 4FGL J0138.6+2923: The optical spectra are available for this source at LAMOST, where it is classified as a QSO at $z = 1.7$. Our classification and redshift are in agreement with these values, as we classify this source as an FSRQ at $z = 1.70$.
    \item 4FGL J0304.5+3349: Due to its strong emission lines, we classify this source as an FSRQ at $z = 0.68$.
    \item 4FGL J0328.9+3514: LAMOST has an optical spectrum where this source is classified as a QSO with a redshift of $z = 0.5$. That value is compatible with the redshift reported here; we classify the source as an FSRQ at $z = 0.50$.
    \item 4FGL J0429.8+2843: We do not distinguish any emission and/or absorption features, thus this source is a bl lac at an unknown redshift.
    \item 4FGL J0436.2$-$0038: Note this is the only source in our sample showing a red spectral continuum. Several explanations could account for this behaviour.  bl lacs are typically found in early-type galaxies, and one possibility for such a red shape is dust absorption in the host galaxy.  However, this seems unlikely, as the spectra show no absorption features, such as the Ca II doublet. Similarly, interstellar dust along the line of sight could cause reddening, but this scenario is also improbable for the same reason. the absence of absorption features in the spectrum.
    
  A more plausible explanation is related to the object’s classification as LSP in the 4LAC catalog, as noted in Table 1. The location of the synchrotron peak in the SED determines the shape of the optical spectrum, as this is often part of the synchrotron emission from the jet in a bl lac \citep[see e.g.][]{1998A&A...333..452K,2011MNRAS.417.1881R}. For this particular source, the  synchrotron peak value reported in the 4LAC is $6.92 \times 10^{13}$ Hz, which lies in the near-IR band. As a result, the optical emission is located on the declining tail of the synchrotron spectrum, giving the continuum a redder appearance. 

    \item 4FGL J0539.7$-$0521c: Due to its strong emission features, we classify this source as an FSRQ at $z = 0.48$.
    \item 4FGL J0614.8+6136: There are strong emission lines, thus this source is an FSRQ at $z = 1.84$.
    \item 4FGL J0704.7+4508: It is a featureless bl lac, and we were not able to determine its redshift. This is our second observation of this object during our BCU optical spectroscopic campaign, since we reported it in Paper I, again finding no emission and/or absorption features.
    \item 4FGL J0713.0+5738: Its strong emission lines allow a FSRQ categorization at a redshift $z =$ 0.82.
    \item 4FGL J0733.0+4915: There were no emission and/or absorption lines in the spectra, thus we classify this source as a bl lac at an unknown redshift.
    \item 4FGL J0746.5$-$0719: This source was observed during our spectroscopic campaign in Paper I, classifying it as an FSRQ due to the Mg II emission feature. However, in this observation cycle, that feature is indistinguishable, and we classify this source as a bl lac. It is a changing-look blazar.
    \item 4FGL J1054.5+2211: 
    This source is not a BCU but is included in our sample due to  inconsistencies in the literature. The redshift reported in the 4LAC catalogue is $z =$ 2.055, and it is classified as a bl lac; however, we have not been able to find the optical spectra that confirm this value. The first available spectra for this source appear in \cite{Shaw2009}, who gave a lower limit of $z >$ 0.60. It was reported in the BZCAT v0.5 catalogue \citep{bzcat}, which classified it as an FSRQ at $z =$ 1.363, but did not disclose the optical spectra for this classification. \cite{Plotkin2010} observed it, giving no spectroscopic redshift value due to its featureless nature. Again, \cite{Shaw2013} observed this source and classified it as a bl lac, providing no value for the spectroscopic redshift due to the lack of emission and/or absorption lines. \cite{pena2021} later observed it and also classified it as a bl lac with unknown redshift.  We do not find any emission and/or absorption features and classify it as a bl lac at an unknown redshift.
    \item 4FGL J1229.1+5521: This source presents strong emission lines, allowing us to measure a redshift of $z = 1.40$ and classify it as an FSRQ.
    \item 4FGL J1517.0+2639: We see a weak absorption feature that we identify as the Ca II H\&K doublet, measuring a redshift of $z = 0.05$. However, this is a very noisy spectrum (S/N = 4) and the absorption could be due to noise, thus this value should be taken as tentative, and we classify it as a bl lac.
    \item 4FGL J1556.6+1417: The optical spectrum of its counterpart shows broad emission lines, thus the classification is an FSRQ at $z = 1.41$.
    \item 4FGL J1754.7+3444: We find four strong emission lines that allow us to classify this source as an FSRQ at $z = 0.62$.
    \item 4FGL J1959.0+3844: The optical spectrum of its counterpart shows broad emission lines; therefore, we classify it as an FSRQ at $z =$ 0.35.
    \item 4FGL J2000.0+4214: There is a strong C III] emission line; this source is classified as an FSRQ at $z = 2.04$.
    \item 4FGL J2120.7+4428: There is a strong Mg II emission line; we classify this source as an FSRQ at $z = 0.59$.
    \item 4FGL J2139.1+3724: There is only a single visible line, which we tentatively identify as Mg II $\lambda$2798, resulting in a redshift lower limit of $z \geq 0.86$. This identification is based on the assumption that the line is Mg II, as it is often the most prominent emission feature in FSRQs and yields the lowest redshift among other prominent lines.
    \item 4FGL J2317.0+3756: We classify this source as an FSRQ at $z = 0.90$ due to its strong emission features.
\end{itemize}

\section{Summary and conclusions}
In this study, we discussed the results of optical spectroscopic observations of 22 sources, 21 BCUs and one bl lac, with the Telescopio Nazionale Galileo (TNG) during the 2022B and 2023A observing cycles,  totaling seven nights. Our analysis revealed that out of the 21 BCUs observed, 14 are classified as FSRQs, indicating the robustness of our selection criteria for potential FSRQs among BCUs. We estimated redshifts for 13 of these FSRQs, and a lower limit for 4FGL J2139.1+3724  due to a single emission line. Remarkably, four FSRQs have redshifts higher than 1, including 4FGL J2000.0+4214 at $z = 2.04$. Six sources were classified as bl lacs, but only one, 4FGL J1517.0+2639, had a tentatively calculated redshift of $z = 0.05$ due to the lack of distinctive spectral features.

The categorization of 4FGL J0128.2+4400 remains inconclusive due to the noisy spectrum (S/N = 2), retaining its BCU status. We re-observed 4FGL J0704.7+4508, previously classified as a featureless bl lac with unknown redshift, and found no new features to determine its redshift. For 4FGL J1754.7+3444, we identified four strong emission lines, reclassifying it as an FSRQ at $z = 0.62$, contrary to its previous classification at $z = 0.016281$. Interestingly, we re-observed 4FGL J0746.5-0719, which showed a featureless spectrum in contrast to previous observations identifying a broad Mg II emission line, suggesting this source could be a changing-look blazar. Additionally, for 4FGL J1054.5+2211, despite literature inconsistencies and previous featureless spectra, our observations also resulted in a featureless bl lac, unable to determine its redshift. Finally, checking the LAMOST LRS General catalogue, we found consistent redshift classifications for 4FGL J0138.6+2923 and 4FGL J0328.9+3514 with our findings, both classified as QSOs.
Overall, our observations and analyses provide insights into the classification and redshift estimation of BCUs, reinforcing the need for continued spectroscopic monitoring to resolve ambiguities in their nature.

\begin{acknowledgements}
We are grateful to the anonymous referee for  constructive comments that have been helpful in improving our manuscript.
This work is partly supported by grant PID2022-138621NB-I00 funded by MCIN/AEI/10.13039/501100011033 and “ERDF A way of making Europe”. A.D. is thankful for the support of Proyecto PID2021-126536OA-I00 funded by MCIN / AEI / 10.13039/501100011033.
Based on observations made with the Telescopio Nazionale Galileo (TNG), located in the Observatorio del Roque de los Muchaschos of the Instituto de Astrofísica de Canarias, operated by the Fundación Galileo Galilei of the Istituto Nazionale di Astrofisica (INAF). This research has made use of ESASky, developed by the ESAC Science Data Centre (ESDC) team and maintained alongside other ESA science mission's archives at ESA's European Space Astronomy Centre (ESAC, Madrid, Spain).
\end{acknowledgements}

\section*{Data Availability}
The data reported in this work are publicly available at the archival database of the TNG telescope.



\bibliographystyle{aa}




\onecolumn
\begin{appendix}
\section{Tables and figures}
\begin{table}
\centering
\small
\setlength\tabcolsep{0pt} 
\caption{List of 4FGL associated sources observed. }
\label{table:listsour} 
\begin{tabular*}{\textwidth}{ l @{\extracolsep{\fill}} *{7}{c} }  
\hline \hline
4FGL Source Name & Association Name & WISE Name & RA  & DEC & Fermi Class &  SED Class    \\
 & & & hms  & dms &  &     \\
\hline
J0128.2+4400 & MG4 J012818+4405 &  J012826.02+440430.6 & 01:28:26 & 44:04:30 & bcu & LSP \\
J0138.6+2923 & B2 0135+291 & 
J013835.32+292204.8 & 01:38:35 & 29:22:05 & bcu & \\
J0304.5+3349 & 4C 33.06 & 
J013835.32+292204.8. & 03:04:41 & 33:48:43 & bcu & LSP \\
J0328.9+3514 & B2 0326+34 &  J032915.35+351005.9 & 03:29:15 &  35:10:06 & bcu & \\
J0429.8+2843 & MG2 J042948+2843 & 
J042949.97+284253.1 & 04:29:50 & 28:42:53 & bcu & LSP \\
J0436.2$-$0038 & NVSS J043614-003637 &  J043614.57-003638.6 & 04:36:15 & -00:36:39 & bcu & LSP \\
J0539.7$-$0521c & TXS 0537-052 &  J053959.93-
051441.2 & 05:40:00 & -05:14:41 & bcu & \\
J0614.8+6136 & GB6 J0614+6139 & 
J061442.16+613908.2 & 06:14:42 & 61:39:08 & bcu & \\
J0704.7+4508 & B3 0701+451 & 
J070450.96+450241.7 & 07:04:51 & 45:02:42 & bcu & LSP \\
J0713.0+5738 & GB6 J0713+5738 & 
J071304.54+573810.2 & 07:13:05 & 57:38:10 & bcu & \\
J0733.0+4915 & TXS 0729+493 & 
J073258.37+491657.5 & 07:32:58 & 49:16:57 & bcu & \\
J0746.5-0710 & PMN J0746-0709 &  J074627.49-
070949.7 & 07:46:27 & -07:09:50 & bcu & \\
J1054.5+2211 & 87GB 105148.6+222705 & J105430.62+221054.9 &	10:54:31 & 22:10:55 & bll & ISP \\
J1229.1+5521 & GB6 J1229+5522 & 
J122909.29+552230.6 & 12:29:09 & 55:22:31 & bcu & \\
J1517.0+2639 & SDSS J151702.59+263858.7 & 
J151702.58+263858.9 & 15:17:03 & 26:38:59 & bcu & LSP \\ 
J1556.6+1417 & TXS 1554+144 &  J155645.56+141549.1 & 15:56:46 & 14:15:49 & bcu & LSP \\
J1754.7+3444  & MG2 J175448+3442 & 
J175451.10+344246.8 & 17:54:51 & 34:42:47 & bcu & \\  
J1959.0+3844 & LQAC 299+038 & 
J195922.01+384654.3 & 19:59:22 & 38:46:54 & bcu & \\
J2000.0+4214 & MG4 J195957+4213 & 
J195958.76+421346.5 & 19:59:59 &	42:13:47 & bcu &  \\
J2120.7+4428 & B3 2118+443 &  J212031.77+443434.3 & 21:20:32 & 44:34:34 	& bcu & 	\\	
J2139.1+3724 & MG3 J213937+3727 & 
J213940.71+372610.6 & 21:39:41 & 37:26:10 & bcu & \\
J2317.0+3756 & B3 2314+377 & 
J231710.28+375948.2 & 23:17:10 & 37:59:48 &  bcu & LSP \\
\hline
\end{tabular*} 
\tablefoot{
Column information are as follows: (1) 4FGL source name; (2) 4FGL associated source name; (3) RA (J2000); (4) Dec. (J2000); (5) 4FGL source class; (6)  4LAC SED class.
}
\end{table}

\begin{table}
\centering
\small
\setlength\tabcolsep{0pt} 
\caption{List of 4FGL associated sources observing information. }
\label{table:listobs} 
\begin{tabular*}{\textwidth}{ l @{\extracolsep{\fill}} *{8}{c} }  
\hline \hline
4FGL Source Name & Obs. Date & R mag   & Exp. time & E(B-V) & SNR & Airmass & Seeing \\
&  yyyy-mm-dd & & s & & & & " \\
\hline
J0128.2+4400 & 2023-01-26 & 20.2 & 5400 & 0.0651 & 2 & 1.37 & 1.2 \\
J0138.6+2923 & 2022-11-20 & 19.8 & 5400 & 0.0425 & 8 & 1.07 & 3 \\
J0304.5+3349 & 2022-11-19 & 18.9 &  3600 & 0.3007 & 10 & 1.01 & 1.3  \\
J0328.9+3514 & 2023-01-25 & 17.3 & 1800 & 	0.2200 & 66 & 1.17 & 1.3 \\
J0429.8+2843 & 2022-11-19 & 20.8 & 7200 & 0.7408 & 8 & 1.13 & 1.6 \\
J0436.2$-$0038 & 2023-01-26 & 18.4 & 3600 & 0.0375 & 15 & 1.36 & 1.2 \\
J0539.7$-$0521c & 2023-01-25 & 18.7 & 3600 & 0.1784 & 30 & 1.26 & 1.22 \\
J0614.8+6136 & 2023-01-24 & 19.9 & 5400 & 0.1909 & 12 & 1.19 & 1.3 \\
J0704.7+4508 & 2023-01-26 & 18.2 & 2400 & 0.0881 & 25 & 1.28 & 1.2 \\
J0713.0+5738 & 2023-01-25 & 19.0 & 2400 & 	0.0461 & 7 &  1.15 & 1.8 \\
J0733.0+4915 & 2023-01-24 & 19.6 & 3600 & 0.0747 & 10 & 1.44 & 1.6 \\
J0746.5$-$0710 & 2023-01-25 & 18.9 & 2400 & 	0.0957 & 6 & 1.54 & 1.8 \\
J1054.5+2211 & 2023-01-24 & 17.2 & 3600 & 0.0138 & 93 & 1.08 & 1.6 \\
J1229.1+5521 & 2023-01-26 & 17.9 & 2400 & 0.0110 & 72 & 1.12 & 1.2 \\
J1517.0+2639 & 2023-06-22 & 19.5 & 3600 & 0.0297 & 4 & 1.01 & 0.7 \\
J1556.6+1417 & 2023-06-23 & 20.2 & 2400 & 0.0388 & 7 & 1.06 & 0.7 \\
J1754.7+3444  & 2023-06-23 & 19.6 & 1800  & 0.0321 & 10 & 1.02 & 0.7 \\
J1959.0+3844 & 2023-06-22, 23 & 20.2  & 3000 & 0.5232 & 5 & 1.09, 1.07 & 0.5, 0.6  \\
J2000.0+4214 & 2023-06-22 & 19.4 & 5400 & 0.3741 & 20 & 1.20 & 0.6 \\
J2120.7+4428 & 2023-06-22 & 21.7 	& 3600 & 0.5558 & 3  & 1.13 & 0.8 \\
J2139.1+3724 & 2022-11-20 & 20.1 & 5400 & 0.0496 & 7 & 1.07 & 2 \\
J2317.0+3756  & 2022-11-19 & 19.9 & 3600 & 	0.1340 & 4 & 1.11 & 1.6  \\
\hline
\end{tabular*}
\tablefoot{
Column information are as follows: (1) 4FGL source name; (2) date of observation (yyyy-mm-dd); (3) apparent R-band magnitude; (4) exposure time in seconds; (5) reddening E(B - V) in mag; (6) mean signal-to-noise ratio of the spectrum; (7) airmass and (8) mean seeing of the exposure in arcsec.
}
\end{table}
\FloatBarrier
\begin{table}
\centering
\small
\setlength\tabcolsep{0pt} 
\caption{List of results. }
\label{table:results} 
\begin{tabular*}{\textwidth}{ l @{\extracolsep{\fill}} *{7}{c} }  
\hline \hline
4FGL Source Name & Observed Line & EW  & Spectral line ID & Detected line type &  z  & Classification \\
          & \AA &     \AA   &                  & &    & \\
\hline
J0128.2+4400 &  & &  &  & ? & BCU \\
J0138.6+2923 & 4070.8 & 89.7 & C IV $\lambda$1549 & emission & 1.70 & FSRQ   \\
                & 4366.8 & 3.5 & He II $\lambda$1640 & emission &  &   \\
                & 5024.1 & 6.4 & Al III $\lambda$1857 & emission &  &  \\
                & 5164.1 & 70.6 & C III] $\lambda$1909 & emission &  &     \\ 
J0304.5+3349 & 4687.0 & 52.0 & Mg II $\lambda$2798 & emission &  0.68 & FSRQ     \\ 
                & 7328.0 & 60.2 & H$\gamma$ $\lambda$4342 & emission & & \\

J0328.9+3514 & 4024.4 & 2.6 & O III] $\lambda$2672 & emission & 0.50
 & FSRQ \\
            & 4119.0 & 21.3 & Mg II $\lambda$2798 & emission &  &     \\ 
            & 5809.4 & 2.3 & [Ne III] $\lambda$3968 & emission & & \\
            & 5840.7& 1.5 & He I $\lambda$3889 & emission & & \\
            & 5959.0 & 6.7 & H$\epsilon$ $\lambda$3971 & emission & & \\ 
            & 6162.0 & 15.8 & H$\delta$ $\lambda$4102 & emission & & \\
            & 6519.1 & 29.2 & H$\gamma$ $\lambda$4342 & emission & & \\
            & 7307 & 66.2  & H$\beta$ $\lambda$4863 & emission & &  \\
            & 7448.5 & 6.5 & O III] $\lambda$4959 & emission & & \\
            & 7521.6 & 21.6 & O III] $\lambda$5007 & emission & & \\
J0429.8+2843 &  & &  &  & ? & bl lac \\
J0436.2$-$0038 &  & &  &  & ? & bl lac \\
J0539.7$-$0521c & 4154.7 & 23.4 & Mg II $\lambda$2798 & emission & 0.48 &  FSRQ   \\ 
            & 6235.0 & 3.4 & H$\delta$ $\lambda$4102 & emission & & \\
            & 7098.0 & 3.4  & H$\beta$ $\lambda$4863 & emission & &  \\
J0614.8+6136 & 5282.6 & 2.6 & Al III $\lambda$1857 & emission & 1.84 & FSRQ \\
                & 5418.5 & 55.5 & C III] $\lambda$1909 & emission &  &     \\ 

J0704.7+4508 &  & &  &  & ? & bl lac \\
J0713.0+5738 & 5099.0 & 74.6 & Mg II $\lambda$2798 & emission & 0.82 & FSRQ  \\
                  & 6800.0 & 23.3 & [O II] $\lambda$3729 & emission &  &  \\
                  & 7064.1 & 17.2 & [Ne III] $\lambda$3870 & emission &  &  \\
J0733.0+4915   &  & &  &  & ? & bl lac \\               
J0746.5$-$0719 &  & &  &  & ? & bl lac \\
J1054.5+2211 &  & &  &  & ? & bl lac \\
J1229.1+5521 & 3470.2 & 65.18 & C IV $\lambda$1549 & emission & 1.40 & FSRQ   \\
                & 4457.3 & 5.0 & Al III $\lambda$1857 & emission &  &  \\
                & 4586.0 & 27.2 & C III] $\lambda$1909 & emission &  &     \\ 
                & 6730.4 & 16.5 & Mg II $\lambda$2798 & emission &  &     \\ 
J1517.0+2639 & 4172.6 & 3.6 & Ca II H $\lambda$3934* & absorption & 0.05? & bl lac \\
                  & 4129.7 & 3.7 & Ca II K $\lambda$3968* & absorption & &     \\

J1556.6+1417 & 5049.5 & 23.0 & [O II] $\lambda$3729 & emission & 0.35 & FSRQ \\
                  & 5235.3 & 7.3 & He I $\lambda$3890 & emission & & \\
                  & 5425.7 & 5.0 & S II $\lambda$4073 & emission & & \\
                  & 5535.8 & 3.8 & H$\delta$ $\lambda$4342 & emission & & \\
                  & 5872.1 & 3.8 & O III] $\lambda$4364 & emission & & \\
                  & 6714.3 & 7.6 & O III] $\lambda$4959 & emission & & \\
                  & 6779.3& 18.1 & O III] $\lambda$5007 & emission & & \\

J1754.7+3444 & 4508.4 & 25.7 & Mg II $\lambda$2798 & emission & 0.62 & FSRQ \\
                    & 6641.6 & 27.8  & H$\delta$ $\lambda$4102 & emission & &  \\
                  & 7029.9  & 20.6 & H$\gamma$ $\lambda$4342 & emission & &  \\
                  & 7877.6 & 52.5  & H$\beta$ $\lambda$4863 & emission & &  \\

J1959.0+3844  & 4475 & 6.7 & Al III $\lambda$1857 & emission & 1.41 & FSRQ    \\
                    & 4562 & 373.4 & C III] $\lambda$1909 & emission & &     \\
                   & 6792 & 46.0 & Mg II $\lambda$2798 & emission & &     \\

J2000.0+4214 & 5663.0 & 2.7 & Al III $\lambda$1857 & emission & 2.04 & FSRQ   \\
                & 5806.9 & 32.5 & C III] $\lambda$1909 & emission &  &     \\    
                    & 7096.18 & 3.7 & C II] $\lambda$2326 & emission &  &    \\

J2120.7+4428 & 4455.5 & 195.8 & Mg II $\lambda$2798 & emission & 0.59 & FSRQ     \\
                & 6172.7 & 1.1 & Ca II H $\lambda$3934 & absorption & &     \\
                 &  6185.7 &  1.7 & Ca II K $\lambda$3968 & absorption & &     \\
J2139.1+3724 & 5214.3 & 20.0 & Mg II $\lambda$2798 & emission & $\geq$ 0.86 & FSRQ     \\
J2317.0+3756 & 5309.1 & 81.4 & Mg II $\lambda$2798 & emission & 0.90 & FSRQ     \\
            & 3806.3 & 1192.2 & C III] $\lambda$1909 & emission &  &     \\    
            & 4362.0 & 31.7 & C II] $\lambda$2326 & emission &  &    \\
            & 7533.3 & 28.0 & [Ne III] $\lambda$1815 & emission &  &     \\  
            & 7763.6 & 27.1 & [S II] $\lambda$4074 & emission &  &     \\  
\hline
\end{tabular*} 
\tablefoot{
Column information are as follows: (1) 4FGL source name; (2) position of the line identified in \AA\, (3) equivalent width of the line identified in \AA\, (4) identification of the line, (5) detected line type, either emission and/or absorption, (6) redshift and (7) classification of the source.

*Tentative identification due to low S/N spectra.

}
\label{tab:results}
\end{table}
\FloatBarrier
\begin{figure*}
    \includegraphics[width=7cm]{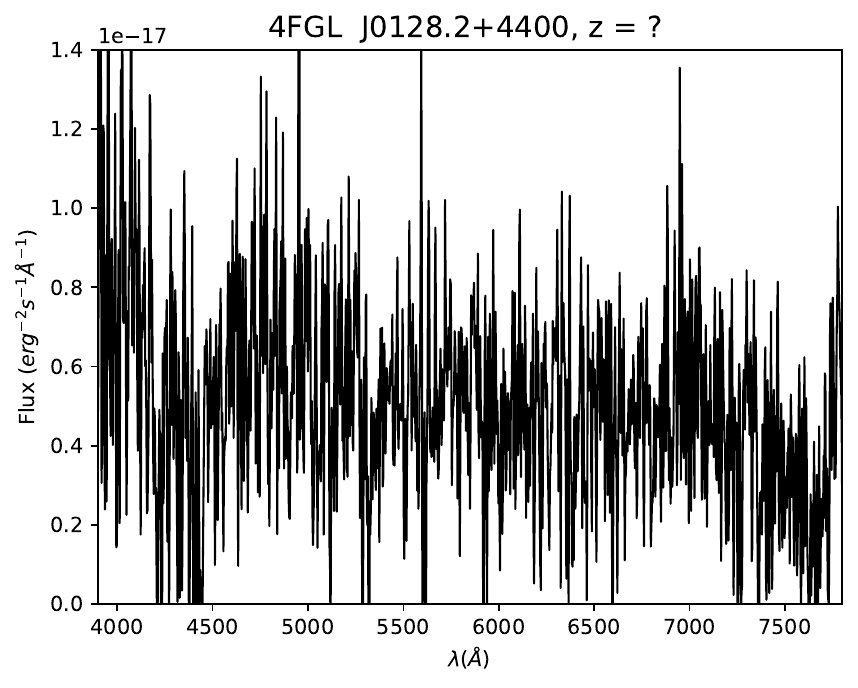}
    \includegraphics[width=7cm]{fc/j0128.pdf}\\
    \includegraphics[width=7cm]{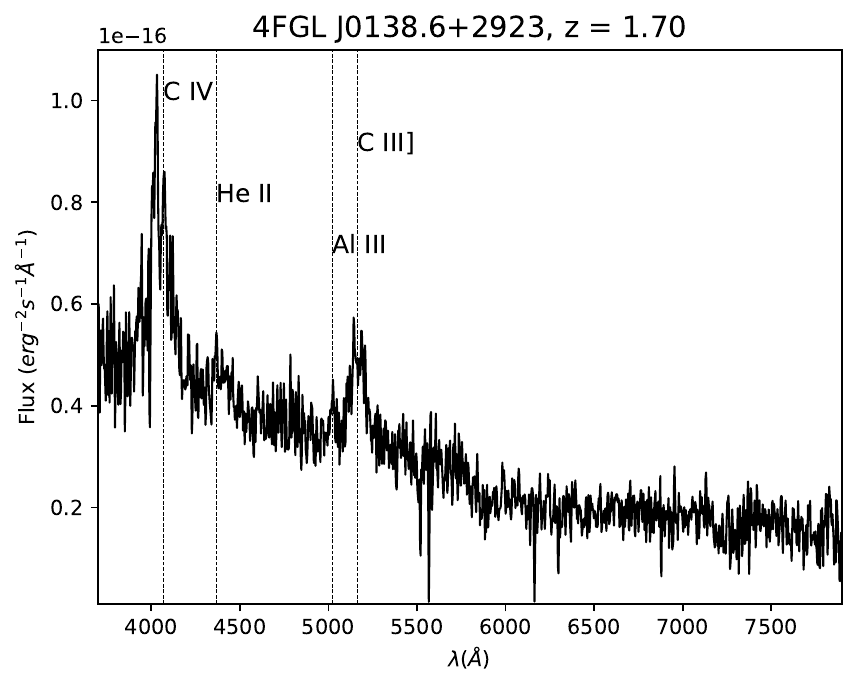}    
    \includegraphics[width=7cm]{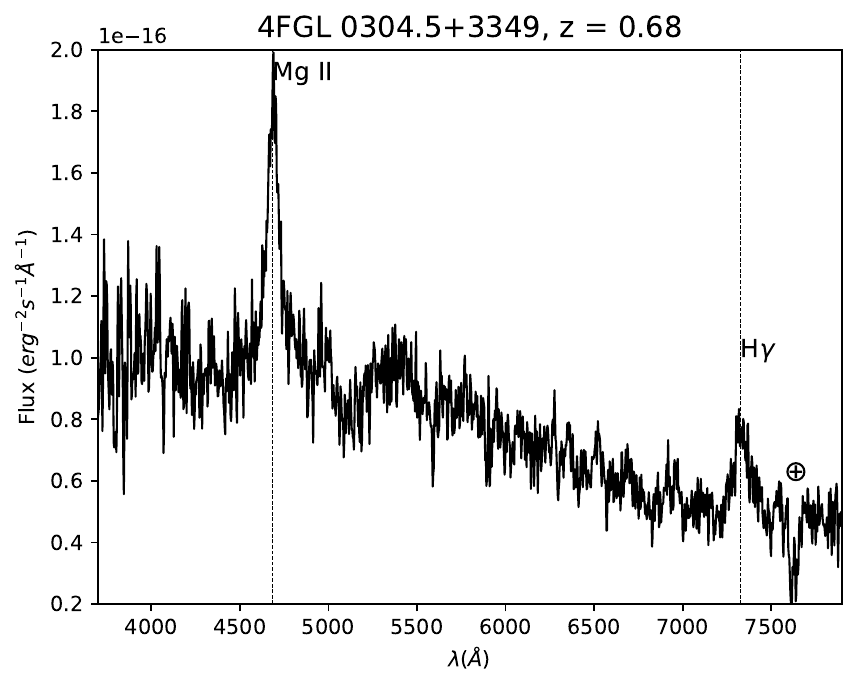}\\
    \includegraphics[width=7cm]{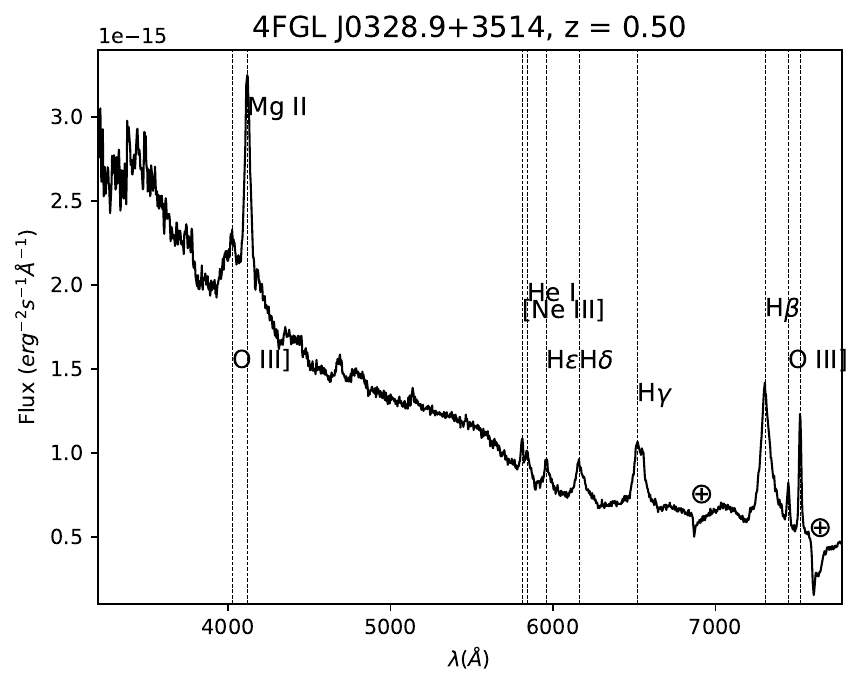}
    \includegraphics[width=7cm]{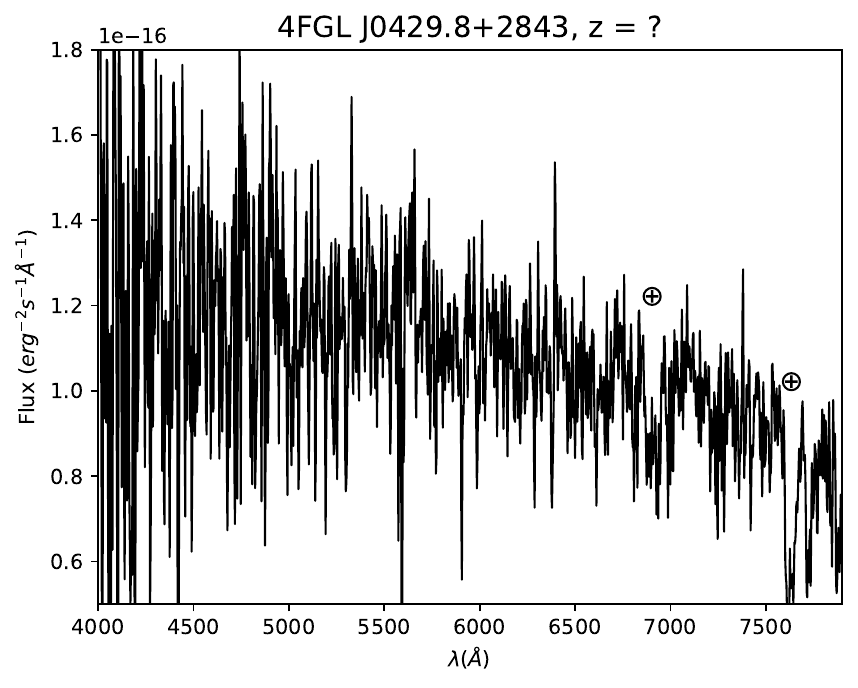}\\
    \includegraphics[width=7cm]{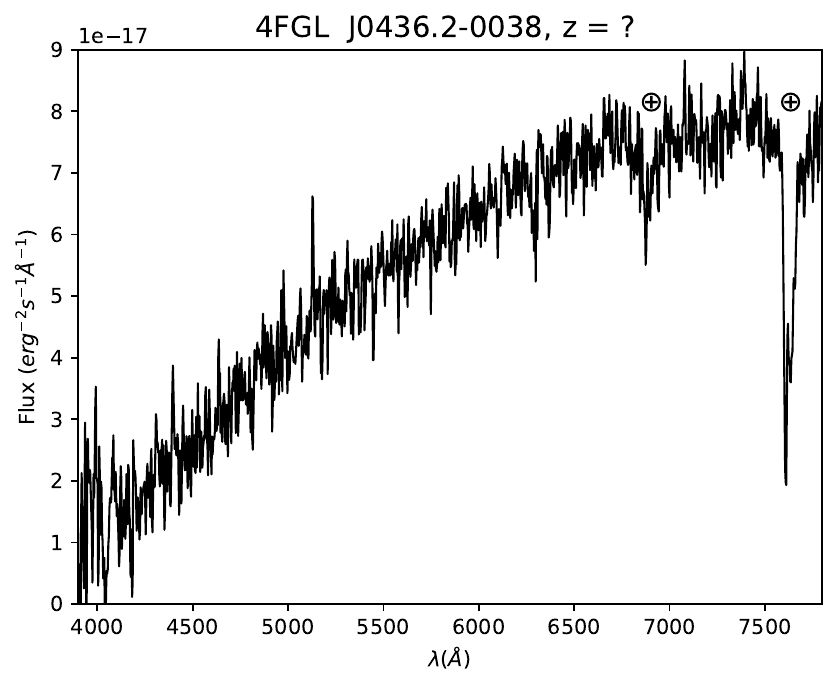}
    \includegraphics[width=7cm]{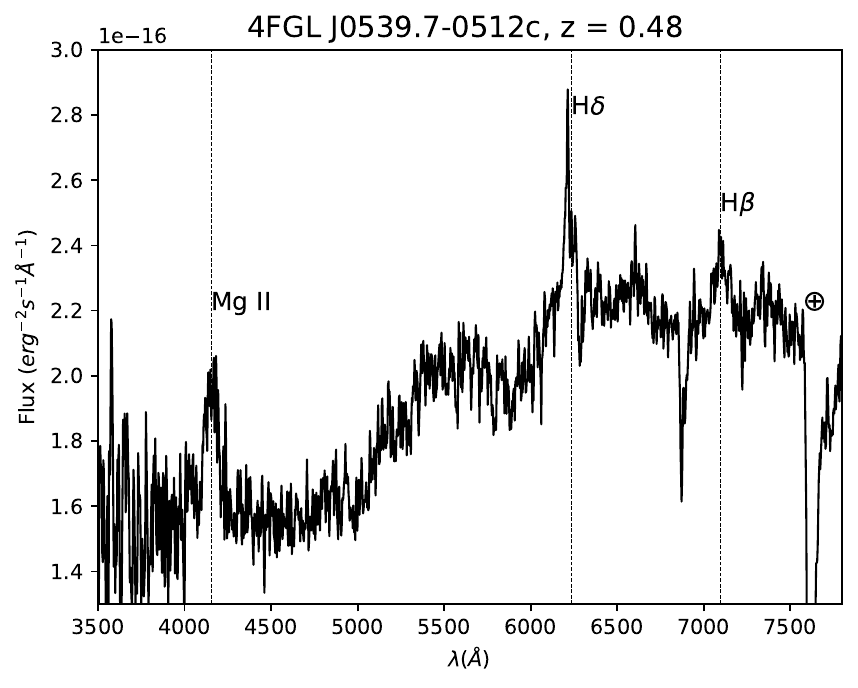}\\
    \label{fig:appendix}
    \caption{Spectra for all sources observed in this work. The name and calculated redshift is provided at the top of each spectra, and observed lines are provided. Crosses indicate telluric absorption lines.  }
\end{figure*}

\begin{figure*}
    \includegraphics[width=7cm]{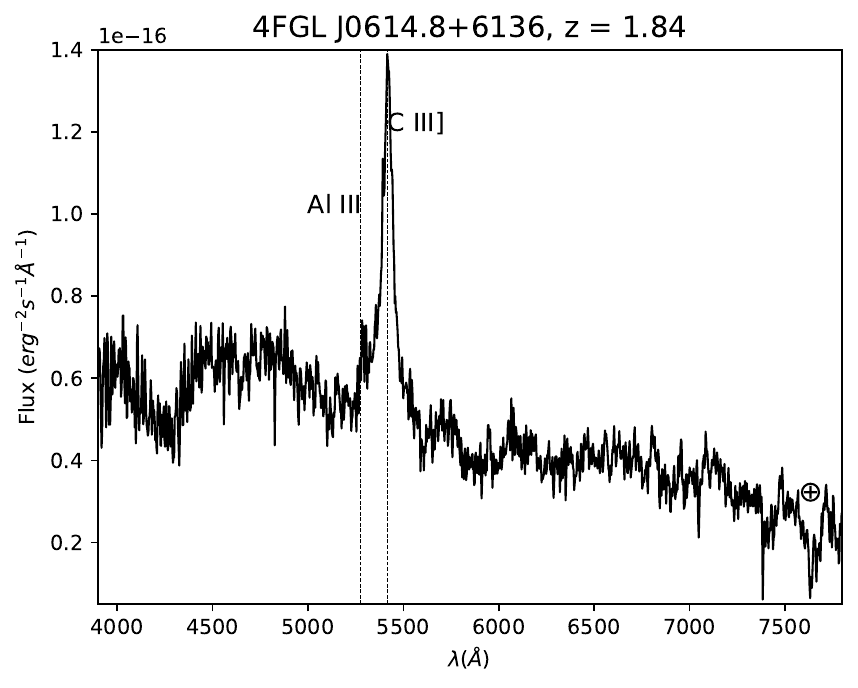}
    \includegraphics[width=7cm]{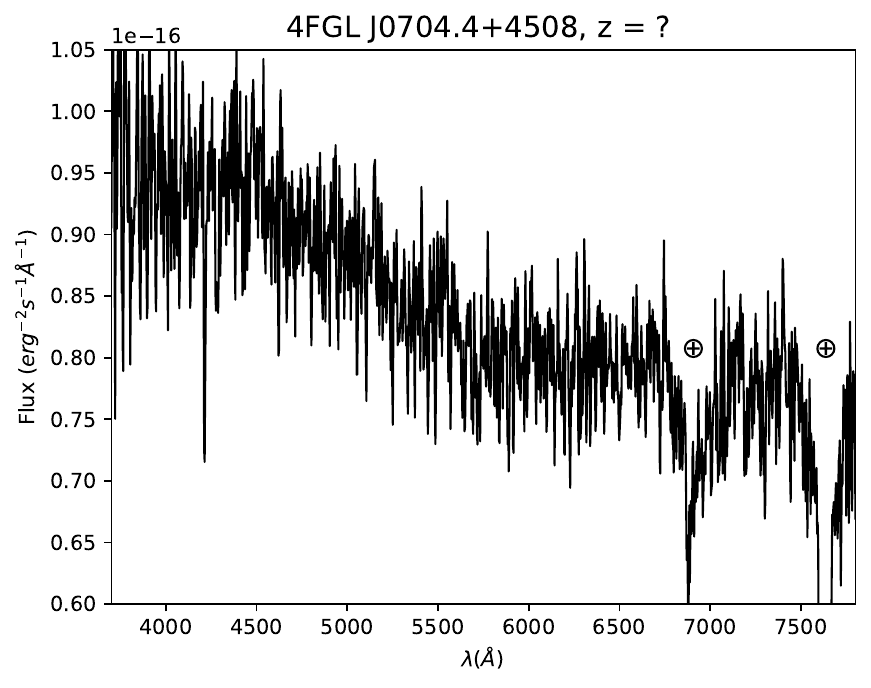}\\
    \includegraphics[width=7cm]{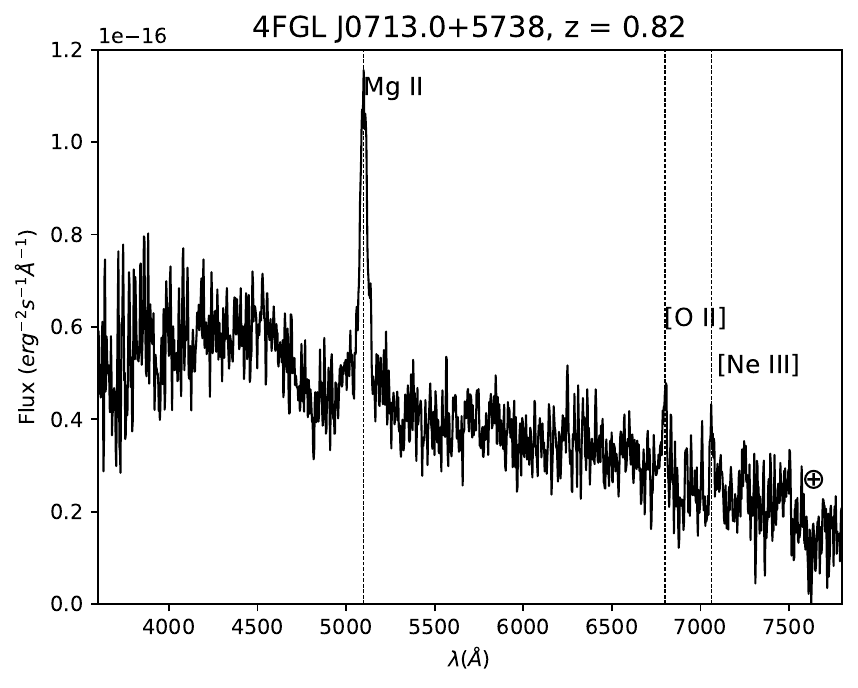}
    \includegraphics[width=7cm]{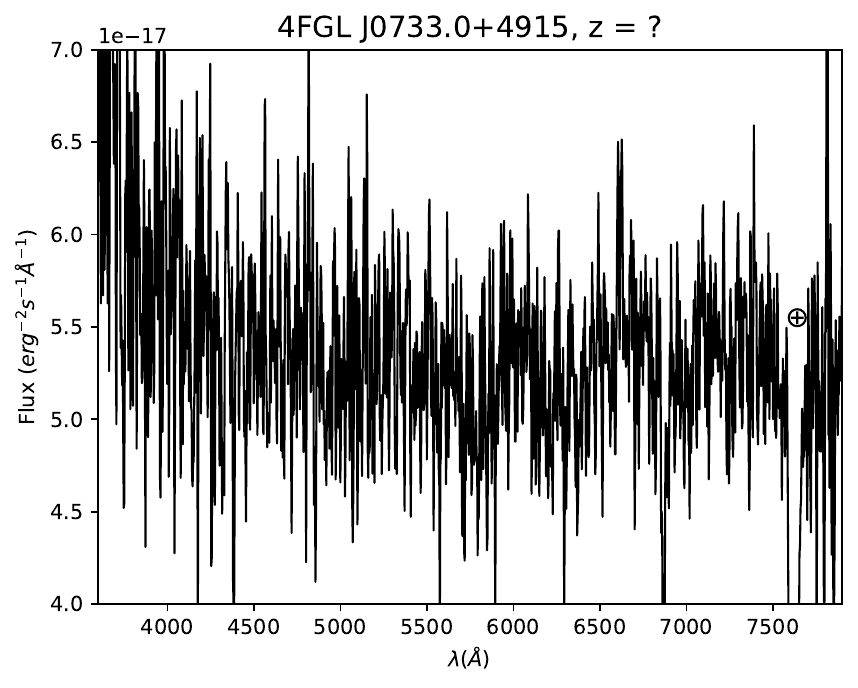}\\
    \includegraphics[width=7cm]{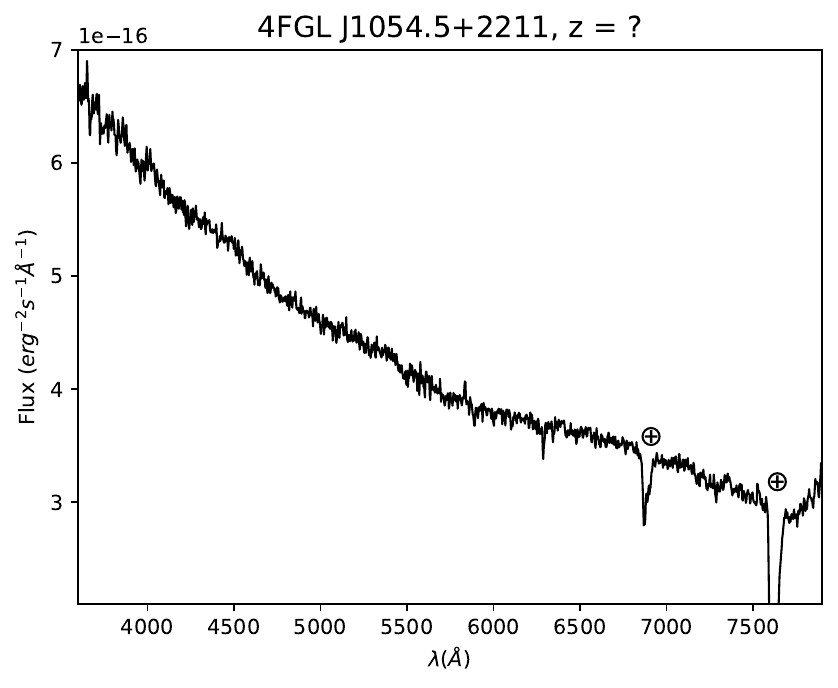}
    \includegraphics[width=7cm]{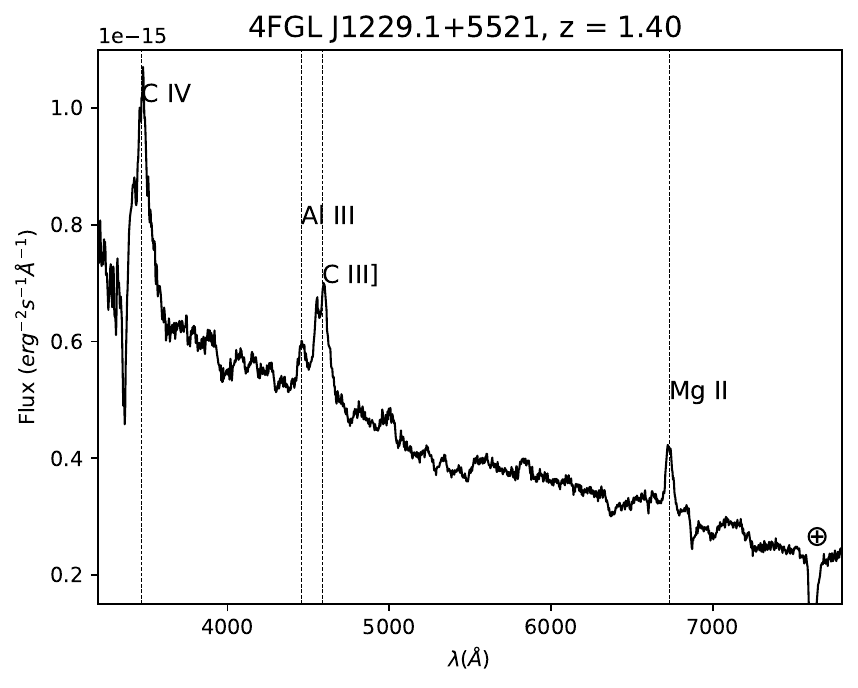}\\
    \includegraphics[width=7cm]{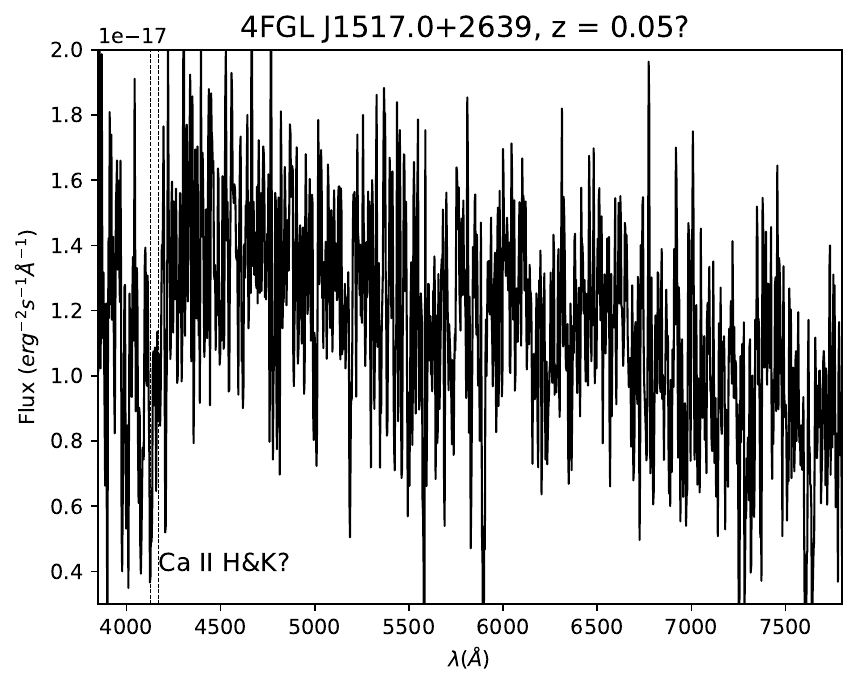}
    \includegraphics[width=7cm]{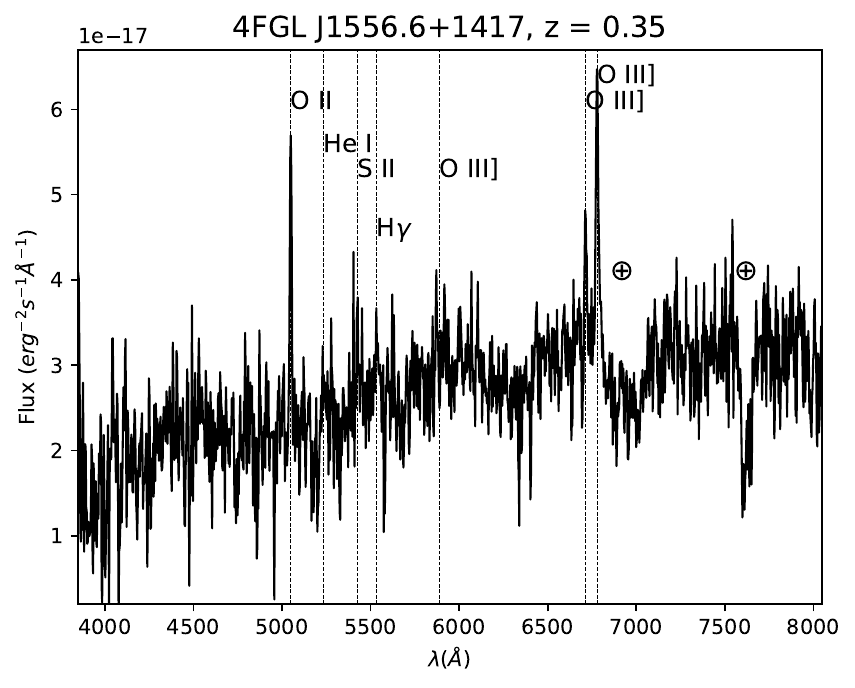}\\
    
\end{figure*}

\begin{figure*}
    
    \includegraphics[width=7cm]{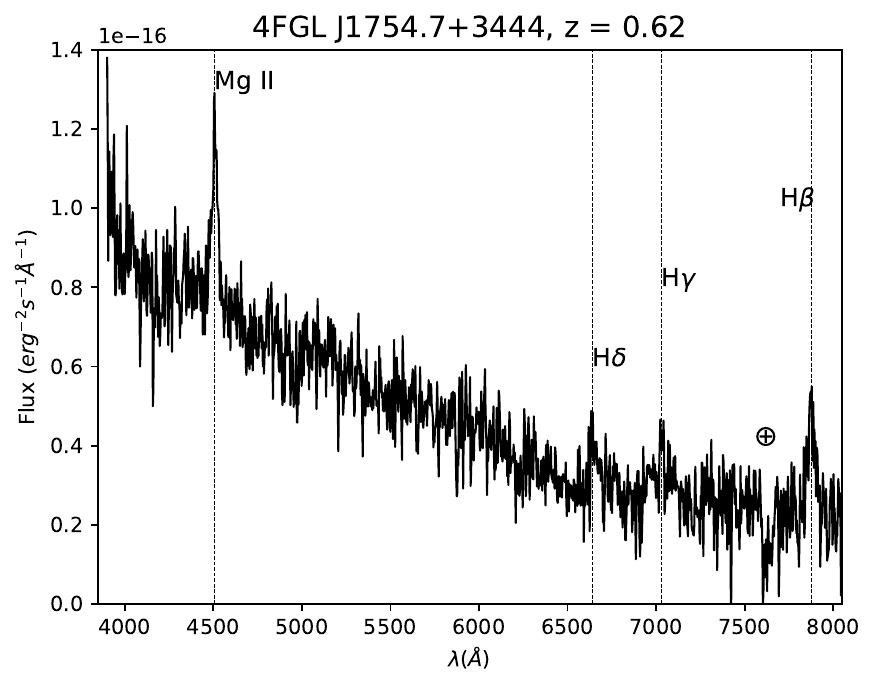}    
    \includegraphics[width=7cm]{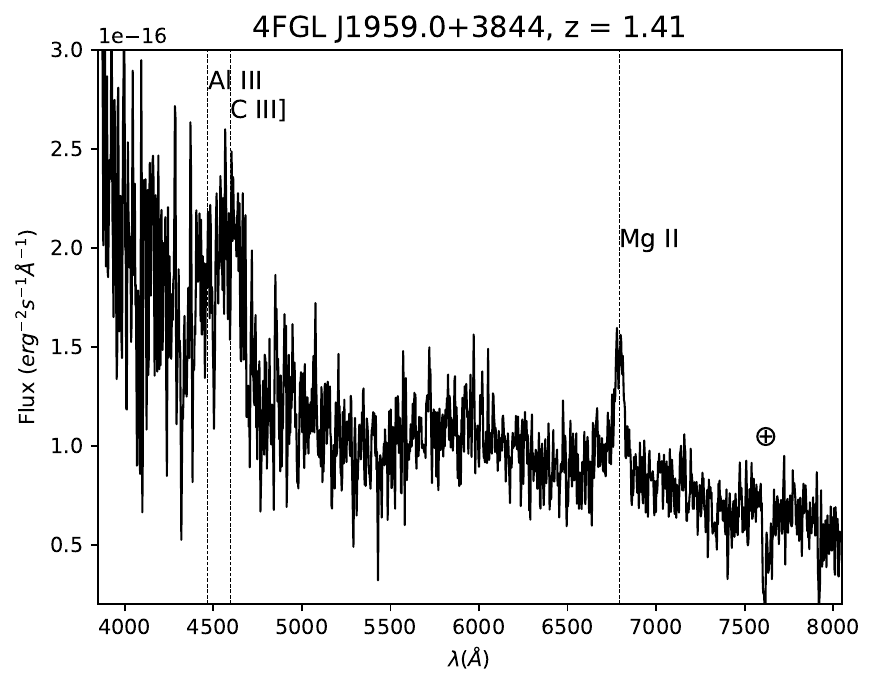}\\
    \includegraphics[width=7cm]{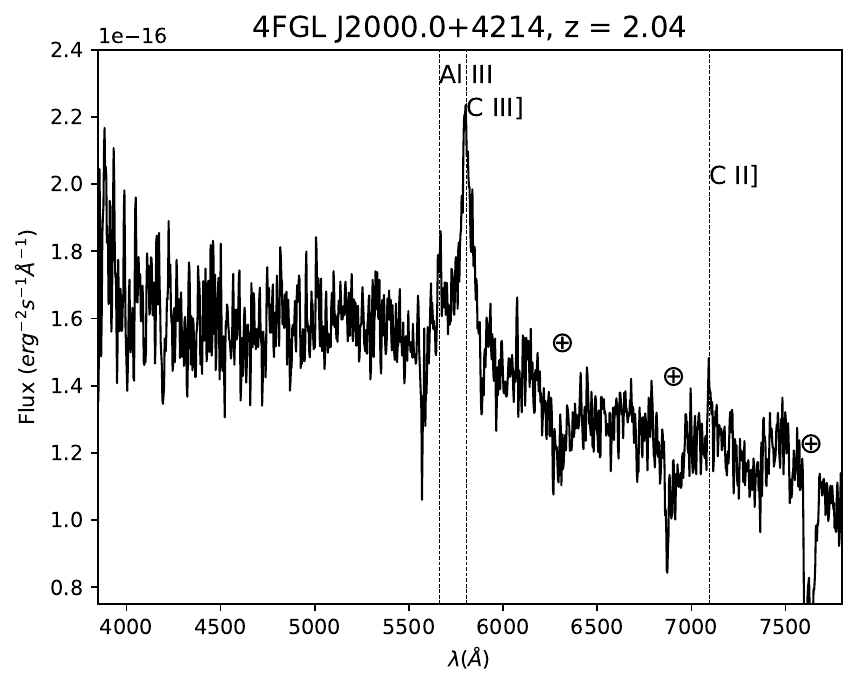}
    \includegraphics[width=7cm]{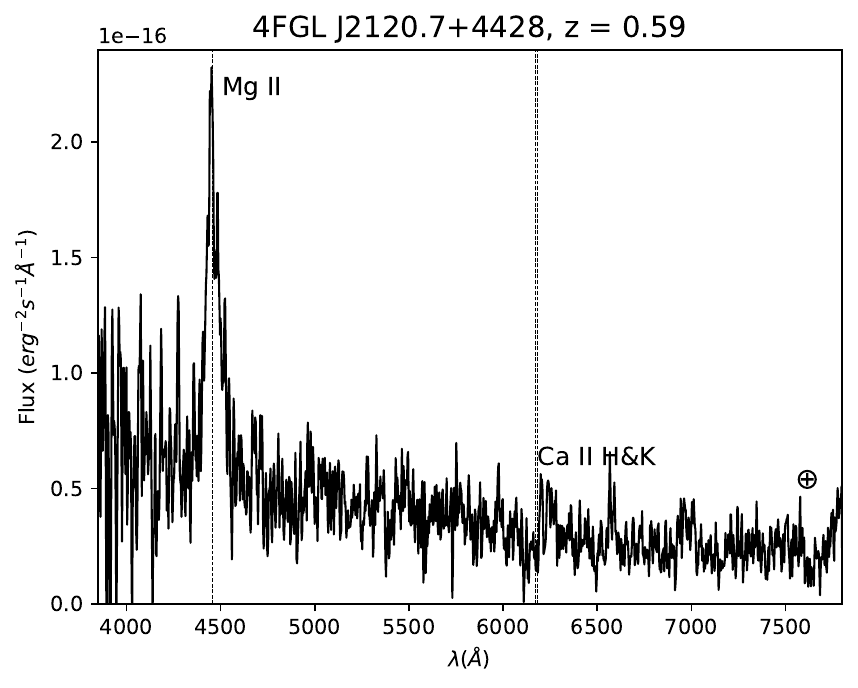}\\
    \includegraphics[width=7cm]{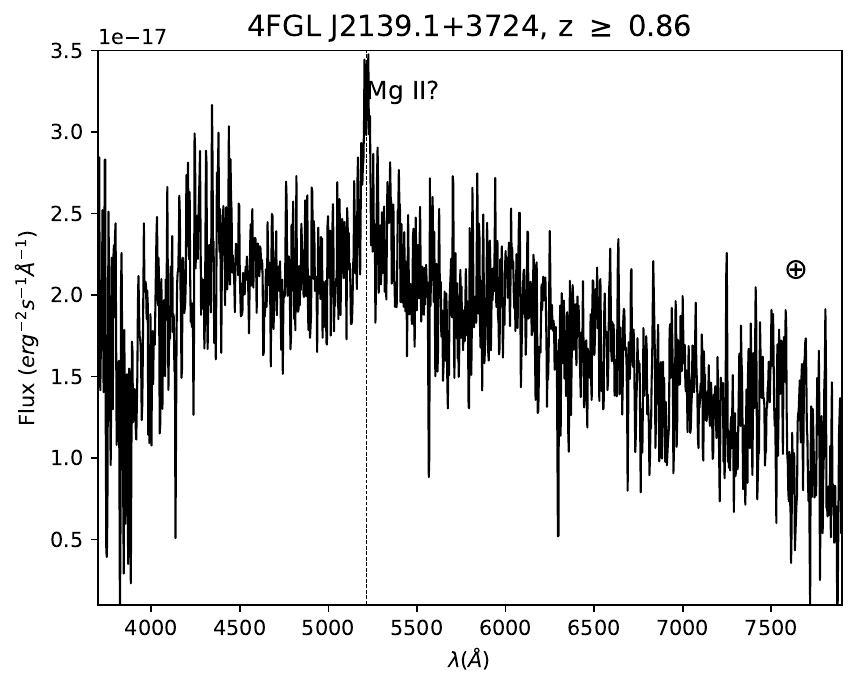}
    \includegraphics[width=7cm]{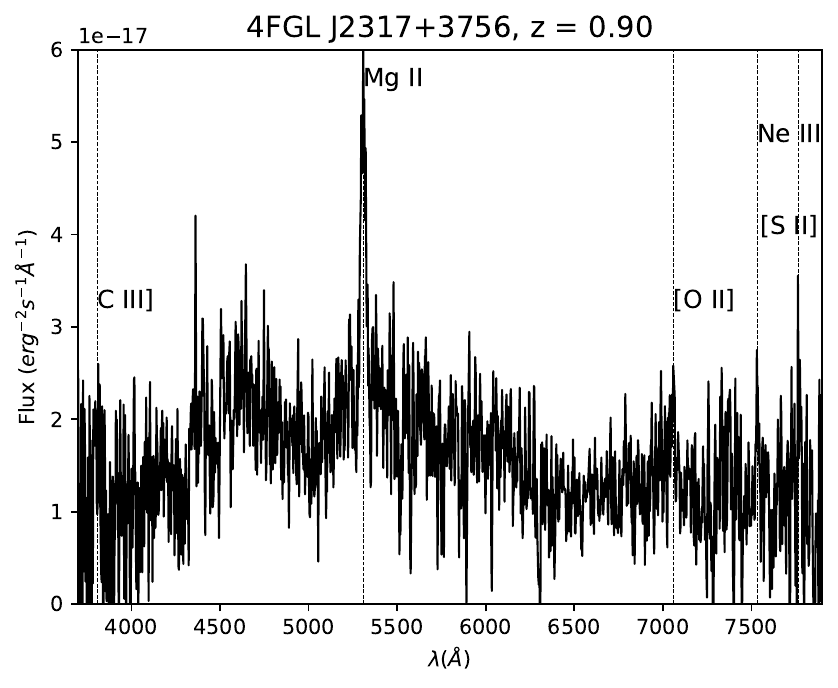}\\    

\end{figure*}


\end{appendix}
\end{document}